\documentclass[12pt]{article}
\input epsf
\usepackage{typearea}
\typearea{12}
\usepackage{amsfonts,amsmath,amssymb}
\usepackage[dvips]{graphicx}
\usepackage{color}

\newcommand{\pa}{\partial}
\newcommand{\nn}{\nonumber}
\newcommand{\Ord}{{\cal O}}
\newcommand{\cA}{{\cal A}}
\newcommand{\ca}{{\mathbf a}}
\renewcommand{\th}{{\Theta}}

\makeatletter

\@addtoreset{equation}{section}
\makeatother

\def\href#1#2{#2}

\begin{document}

\begin{titlepage}
\begin{flushright}
{NCTS-TH/1507}
\end{flushright}

\begin{center}

\hfill 
\vskip 1.2in

\textbf{\Large The IR Obstruction to UV Completion}\\[3mm]
\textbf{\Large for Dante's Inferno Model with}\\[3mm]
\textbf{\Large Higher-Dimensional Gauge Theory Origin}\\[18mm]

{Kazuyuki Furuuchi${\,}^a$ and Yoji Koyama${\,}^b$}
\vskip8mm
${}^a\,${\sl Manipal Centre for Natural Sciences, Manipal University}\\
{\sl Manipal, Karnataka 576104, India}\\[1mm]
${}^b\,${\sl National Center for Theoretical Sciences, 
National Tsing-Hua University}\\
{\sl Hsinchu 30013, Taiwan R.O.C.}

\vskip8mm 
\end{center}
\begin{abstract}
We continue our investigation of
large field inflation models 
obtained from higher-dimensional gauge theories,
initiated in our previous study \cite{Furuuchi:2014cwa}.
We focus on Dante's Inferno model
which was the most preferred model in 
our previous analysis.
We point out the relevance of
the IR obstruction to UV completion, 
which constrains
the form of the potential of the massive vector field,
under the current observational upper bound
on the tensor to scalar ratio.
We also show that in simple examples of
the potential arising from 
DBI action of a D5-brane and that of an NS5-brane
that the inflation takes place in the field range
which is within the convergence radius 
of the Taylor expansion. 
This is in contrast to
the well known examples of axion monodromy inflation
where inflaton takes place outside the convergence radius
of the Taylor expansion.
This difference arises 
from the very essence of Dante's Inferno model
that the effective inflaton potential is
stretched in the inflaton field direction
compared with the potential for the original field.
\end{abstract}

\end{titlepage}

\section{Introduction}\label{secIntro}

Effective field theories\footnote{%
For a review of effective field theory, 
see for example \cite{Georgi:1994qn}.}
allow us to make predictions with desired accuracy 
without knowing the full details of 
the underlying UV theory.
Traditional attitude to
effective field theories was that
all the terms allowed by the symmetries
should appear in the action,
and there is no theoretical constraints on them
if one does not know the underlying UV theory
but one can estimate natural magnitude of their coefficients.
However, this view was challenged by the suggestions that 
some reasonable properties which any UV theory should satisfy
impose certain constraints on effective field theories 
\cite{Vafa:2005ui,ArkaniHamed:2006dz,Adams:2006sv}.
In the context of inflation, 
one of the most studied such criteria is
the weak gravity conjecture \cite{ArkaniHamed:2006dz}.
It states that
in order for an effective field theory 
with a massless Abelian gauge field 
to be consistently coupled to gravity,
there exists at least one charged particle in the spectrum 
to which the gauge force acts stronger than the gravitational force.
The weak gravity conjecture was proposed to explain
why extra-natural inflation \cite{ArkaniHamed:2003wu},
in which a higher-dimensional component of a gauge field
plays the role of inflaton,
appeared to be difficult to realize in string theory.
In the simplest 
single-field
extra-natural inflation model, 
the weak gravity conjecture restricts
the inflaton field range to be sub-Planckian,
making the model observationally unfavored.
The restriction from the weak gravity conjecture
in general multi-axion inflation models
has been a subject of recent extensive studies
\cite{Cheung:2014vva,Cheung:2014ega,Rudelius:2014wla,Bachlechner:2014gfa,%
delaFuente:2014aca,Rudelius:2015xta,Montero:2015ofa,%
Brown:2015iha,Bachlechner:2015qja,Hebecker:2015rya,
Brown:2015lia,Junghans:2015hba,%
Heidenreich:2015wga,Heidenreich:2015nta,%
Kooner:2015rza,Palti:2015xra,Kappl:2015esy}.\footnote{%
Since the weak gravity conjecture is not the main target of
the current article (though it is related in the 
broader perspective of 
constraining effective field theories from UV consistencies),
we did not attempt to make a complete list of references on it here.
We picked up the articles which had attracted our attention
while investigating the main theme of this article.}

In this article,
we would like to examine another\footnote{%
Possible relation between the weak gravity conjecture
and the IR obstruction to UV completion 
has been speculated in \cite{ArkaniHamed:2006dz}.
See \cite{Cheung:2014ega} for an investigation in this direction.}
criterion for effective field theories
to be embedded in a consistent UV theory:
The IR obstruction of UV completion
\cite{Adams:2006sv},
applied to theories with massive vector fields
\cite{Hashimoto:2008tw}.\footnote{%
See \cite{Baumann:2015nta} 
which discusses the analyticity issue in inflation.
Note that our interest is on the
IR obstruction to UV completion for 
effective field theories with massive vector fields
\cite{Hashimoto:2008tw},
which has not been discussed before 
in the context of inflation as far as we have noticed.}
In \cite{Adams:2006sv}, 
it was argued that 
the pathological behavior of an effective field theory,
namely the superluminal 
propagation of
fluctuations
around certain backgrounds,
is closely related to
the obstruction for the effective field theory 
to be embedded in a UV theory whose $S$-matrix
satisfies unitarity and canonical analyticity constraints.
The obstruction to the UV completion
was probed through the analytic 
property of the forward scattering amplitude 
of the effective field theory.
In \cite{Hashimoto:2008tw},
the same type of analyticity property
was used to argue that
a massive vector field theory which has
a Lorentz-symmetry-breaking local minimum
cannot be embedded in UV theories
whose $S$-matrix satisfies unitarity and
canonical analyticity property. 
Incidentally, the constraints 
on the coefficients of the potential
of the massive vector field
found in \cite{Hashimoto:2008tw}
were the same as the constraints 
derived by requiring causal propagation
of the massive vector field  
\cite{Velo:1970ur}.
Thus also in the massive vector field theory,
the acausal propagation in the IR 
appears to be the obstruction to UV completion.

In our previous article \cite{Furuuchi:2014cwa},
we surveyed large-field inflation models obtained
from higher-dimensional gauge theories.
We discussed naturalness of the parameter values
allowed by the observational constraints 
together with the theoretical constraints from
the weak gravity conjecture.
We concluded that
Dante's Inferno model was most natural 
among the models studied in \cite{Furuuchi:2014cwa}.
At the time when we were writing \cite{Furuuchi:2014cwa},
BICEP2 had suggested large tensor-to-scalar ratio $r$ \cite{Ade:2014xna},
therefore we took $r = 0.16$ as our reference value.
However, later analysis
indicates that the 
analysis of \cite{Ade:2014xna}
underestimated the contribution from polarized dusts
\cite{Ade:2015tva,Ade:2015xua,Ade:2015lrj}.
These analysis gave lower upper bound on $r$
compared with \cite{Ade:2014xna},
for example $r < 0.12$ at $95\%$ CL in \cite{Ade:2015tva},
which is also consistent with the earlier analysis
\cite{Planck:2013jfk}.\footnote{%
While we were finalizing the current article,
a new tighter bound on the tensor-to-scalar
ratio has been announced 
by Keck Array \& BICEP2 collaborations 
\cite{Array:2015xqh}.
As our analysis had already finished with the earlier bound,
and we would also like to see if the new bound will
be confirmed with other independent experiments,
we will not consider the bound given in \cite{Array:2015xqh}
in this article.}
This updated upper bound on the tensor-to-scalar ratio
does not qualitatively 
change our previous
conclusion that Dante's Inferno model 
is most preferred
in our framework.
However, it does make the 
chaotic inflation with
quadratic potential
which was used in \cite{Furuuchi:2014cwa}
moderately disfavored \cite{Ade:2015lrj}.
To accommodate the updated upper bound
of the tensor-to-scalar ratio,
in this article
we include quartic term to the potential
of massive vector field,
and this is the place
where the IR obstruction to UV completion is relevant:
It constrains the sign of the quartic term in the potential
to be negative (in the convention described in the main text).
We show that this sign is actually favorable 
when comparing the model with the
updated upper bound on the tensor-to-scalar ratio.
These will be discussed in section \ref{secDI}.

In section \ref{secDBI}, we examine DBI action
which was used in the axion monodromy inflation
\cite{McAllister:2008hb}.
DBI action is a low energy effective field theory
of string theory whose $S$-matrices satisfy 
unitarity and canonical analyticity
constraints.
Therefore, it is a good example for
testing whether the arguments of the
IR obstruction to UV completion
\cite{Adams:2006sv} 
were correct.
Indeed in \cite{Adams:2006sv} 
it was shown that 
the embedding fields satisfy the constraints
from IR obstruction to UV completion.
In the current work, we are interested in 
the NS-NS (or RR) two-form field 
in six-dimensional DBI action on 5-branes,
which upon dimensional reduction to five-dimensions
gives a massive vector field.
The five-dimensional model can be treated in
a similar way as in the section \ref{secDI},
but the potential for the massive vector field
contains higher order terms.
One of the main interests here is
the effects of these higher order terms.
Using the parameter values allowed by the CMB data 
obtained in section \ref{secDI},
we show in simple examples that the
inflation takes place in the field range
which is within the convergence radius of
the Taylor expansion of the DBI action.
This means that the linear approximation
of the potential at large field,
which was appropriate in the 
well known examples of
axion monodromy inflation \cite{McAllister:2008hb},
is not valid in Dante's Inferno model,
in the simple models we study.
This difference originates from the very essence
of Dante's Inferno model
that the inflaton potential
is stretched in the inflaton field direction
compared with the potential
of the original field 
due to a field redefinition.

We summarize with discussions on future directions in section \ref{secsd}.

\section{The IR obstruction to UV completion
for Dante's Inferno model with
higher-dimensionsional gauge theory origin}\label{secDI}

Dante's Inferno model \cite{Berg:2009tg}
is a two-axion model described by the following potential
in four dimensions:
\begin{equation}
V_{DI}(A,B) 
= 
V_A(A) 
+ 
\Lambda^4 
\left\{
1 - \cos 
 \left(
 \frac{A}{f_A} - \frac{B}{f_B}
 \right)
\right\} .
\label{VDI}
\end{equation}
The potential 
(\ref{VDI}) appears as a 
leading approximation
to the effective potential obtained from
the following five-dimensional gauge theory
compactified on a circle:\footnote{%
We used the charged fermion as an example of charged matters.
One may consider different matter fields, 
it does not affect the conclusion qualitatively
as long as the charge assignment is similar.}
\begin{align}
S =
\int d^5 x&
\Bigl[
-\frac{1}{4} F_{MN}^{(A)} F^{(A)MN}
- V_A( \cA_{M})
-\frac{1}{4} F_{MN}^{(B)} F^{(B)MN} \nn \\
& \quad 
- i \bar{\psi} \gamma^M \left( \pa_M + i g_{A5} A_M - i g_{B5} B_M \right) \psi 
\Bigr], \nn \\
&\qquad \qquad \qquad (M,N = 0,1,2,3,5) ,
\end{align}
where
\begin{equation}
\cA_M = A_M - g_{A5} \pa_M \theta ,
\label{cA}
\end{equation}
and the field strengths of the Abelian gauge fields are given as
\begin{equation}
F^{(A)}_{MN} = \pa_M A_N - \pa_N A_M, \quad
F^{(B)}_{MN} = \pa_M B_N - \pa_N B_M .
\label{FMN}
\end{equation}
We consider the diagonal kinetic term for the gauge fields
for simplicity.

Since the metric convention will be important 
in the following discussions,
we explicitly state here that our convention is
\begin{equation}
\eta_{MN} = \mbox{diag} (+ - - - - ).
\label{metric}
\end{equation}

The axion decay constants in four-dimension 
are related to parameters in the five-dimensional
gauge theory as
\begin{equation}
{f_{A}} = \frac{1}{g_{A} (2\pi L_5)}, \quad 
{f_{B}} = \frac{1}{g_{B} (2\pi L_5)} ,
\label{fAfB}
\end{equation}
where $L_5$ is the compactification radius of the fifth dimension,
and $g_A$ and $g_B$ are four-dimensional gauge couplings 
which are related to the five-dimensional gauge couplings 
$g_{A5}$ and $g_{B5}$ as
\begin{equation}
g_A = \frac{g_{A5}}{\sqrt{2\pi L_5}}, \quad
g_B = \frac{g_{B5}}{\sqrt{2\pi L_5}} .
\label{gAB}
\end{equation}
We consider the potential of the vector field $\cA_M$ 
given in the power series expansion:
\begin{equation}
V_A(\cA_M) = v_2 \cA_M \cA^M + v_4 (\cA_M \cA^M)^2 
+ v_6 (\cA_M \cA^M)^3
+ \cdots 
= 
\sum_{n=1}^\infty
v_{2n} (\cA_M \cA^M)^{n} .
\label{VAA}
\end{equation}
From the effective field theory point of view,
the functional form of the
potential $V_A(\cA_M)$
is arbitrary as long as it respects Lorentz symmetry,
which is already implemented in (\ref{VAA}).
However, 
it has been claimed that
there are
certain constraints on the potential
in order for the effective field theory
to be derived from a UV theory
whose $S$-matrix satisfies unitarity 
and canonical analyticity constraints  
\cite{Adams:2006sv}.
In the case of massive vector field theories
which is of our current interest,
this issue was taken up by \cite{Hashimoto:2008tw}.
The following sign constraints were derived from
the condition that
the effective field theory to be embedded
to a unitary UV theory with canonical analyticity property:
\begin{equation}
v_2, v_4 < 0.
\label{v4}
\end{equation}
Note that our metric convention (\ref{metric})
follows that in \cite{Hashimoto:2008tw}.
Incidentally, 
(\ref{v4}) is the same condition given in \cite{Velo:1970ur}
for the massive vector field theory
to have causal evolution.
As we are interested in a model which has a sound 
IR behavior as well as an origin in a sane UV theory,
below we assume that (\ref{v4}) is satisfied.

The naturalness in five-dimension dictates
$v_{2n} = c_{2n}/\Lambda_{UV}^{3n-5}$
with $c_{2n} \sim \Ord(1)$,
where $\Lambda_{UV}$ is the UV cut-off scale
at which the effective field theory breaks down,
if there were no symmetry to forbid these terms.
However, if there is an approximate symmetry,
it is natural that
the coefficients of the terms which violate the symmetry
is small \cite{'tHooft:1979bh}.
In the current case,
$|c_{2n}| \ll 1$ 
with
$g_{A5} \sqrt{\Lambda_{UV}}\ll 1$
is natural 
because turning off these couplings
recovers the $U(1)$ gauge symmetry
without the charged matter fields and the Stueckelberg field.

When 
\begin{equation}
\left|
{\cal A}_M {\cal A}^M
\right|
\ll \Lambda_{UV}^3,
\label{conv}
\end{equation}
dropping the terms with $n \geq 3$ 
in (\ref{VAA}) will be a good approximation.
Whether (\ref{conv}) is realized or not
depends both on 
$\left|
{\cal A}_M {\cal A}^M
\right|$
required for inflation which 
we will obtain below,
and on $\Lambda_{UV}$,
which depends on the microscopic
origin of the effective field theory.\footnote{%
One can put the upper bound on $\Lambda_{UV}$ by
estimating the magnitude of loop corrections,
but new physics can enter before the upper bound is reached.}
As the field range of the original fields
are restricted in Dante's Inferno model
as we review below,
it is natural to expect that
(\ref{conv}) would hold in many cases,
but one should examine it
for each microscopic model.
In this section
we assume that (\ref{conv}) is satisfied and
set $v_{2n} = 0$ for $n\geq 3$ in (\ref{VAA}):
\begin{equation}
V_A(\cA_M) = v_2 \cA_M \cA^M + v_4 (\cA_M \cA^M)^2.
\label{V4}
\end{equation}
We will examine 
how good the truncation of the potential 
at the quartic order is
in explicit microscopic models based on 5-branes
in section \ref{secDBI}.

The four-dimensional
effective potential for the zero-modes 
$A$ and $B$ of the fifth components of the gauge fields
$A_5$ and $B_5$, respectively
is given at one-loop order as follows
(the details of the calculations are given in 
appendix \ref{apponeloop}):
\begin{equation}
V_{1-loop}(A,B) = V_{cl}(A) + V_g(A) + V_f(A,B).
\label{Vtot}
\end{equation}
Here, the classical part of the potential, 
\begin{equation}
V_{cl}(A) =\frac{1}{2}m^2A^2 - \frac{\lambda}{4!}A^4 ,
\label{Vcl}
\end{equation}
directly follows from 
the classical potential (\ref{V4}) 
upon dimensional reduction.\footnote{
Remember the metric sign convention in (\ref{metric})! 
} 
In (\ref{Vcl}) 
we introduced parametrization
suitable in four-dimension:
\begin{equation}
- v_2 = \frac{m^2}{2} > 0, \quad 
- \frac{v_4}{2\pi L_5} = \frac{\lambda}{4!} > 0,
\label{v2v4}
\end{equation}
where the sign follows from the constraints from
the IR obstruction to UV completion, Eq.~(\ref{v4}).
As shown in the appendix \ref{apponeloop},
the one-loop contribution from the fermion $V_f(A,B)$
in (\ref{Vtot}) is given as
\begin{equation}
V_f(A,B)
=
\frac{3}{\pi^2 (2 \pi L_5)^4}
\sum_{n=1}^{\infty}\frac{1}{n^5}\cos 
\left\{
n\left(\frac{A}{f_A}-\frac{B}{f_B}\right)
\right\} .
\label{Vf}
\end{equation}
$V_g(A)$ 
in (\ref{Vtot})
is the one-loop contribution from the gauge field $A_M$.
As shown in the appendix \ref{apponeloop},
the contribution of this term 
is sub-leading compared with that 
of the classical potential $V_{cl}(A)$
when 
\begin{equation}
2\pi L_5 \gtrsim 1 \times 10^2 ,
\label{L5lb}
\end{equation}
in the parameter region and the field value of our interest
which are to be discussed below.
Here and below we use the unit $M_P = 1$,
where $M_P$ is the reduced Planck scale
$M_P = (8\pi G_N)^{-1/2} \simeq 2.4 \times 10^{18}$ GeV.
Since the five-dimensional gauge theory
is non-renormalizable and should be regarded as 
an effective field theory,
we do not expect the compactification radius $L_5$
to be too close to the Planck scale.
Therefore (\ref{L5lb}) is a reasonable assumption to make.
Below we adopt this assumption and drop $V_g(A)$ from the analysis below.
However, though this is a reasonable assumption,
it is also for technical simplicity.
Dante's Inferno model
may still work even if the contribution from $V_g(A)$
is not negligible,
though the loop expansion should be under control
for analyzing such case.

\begin{figure}[t!]
 \centering
 \vspace{-0cm}
  \includegraphics[width=12cm]{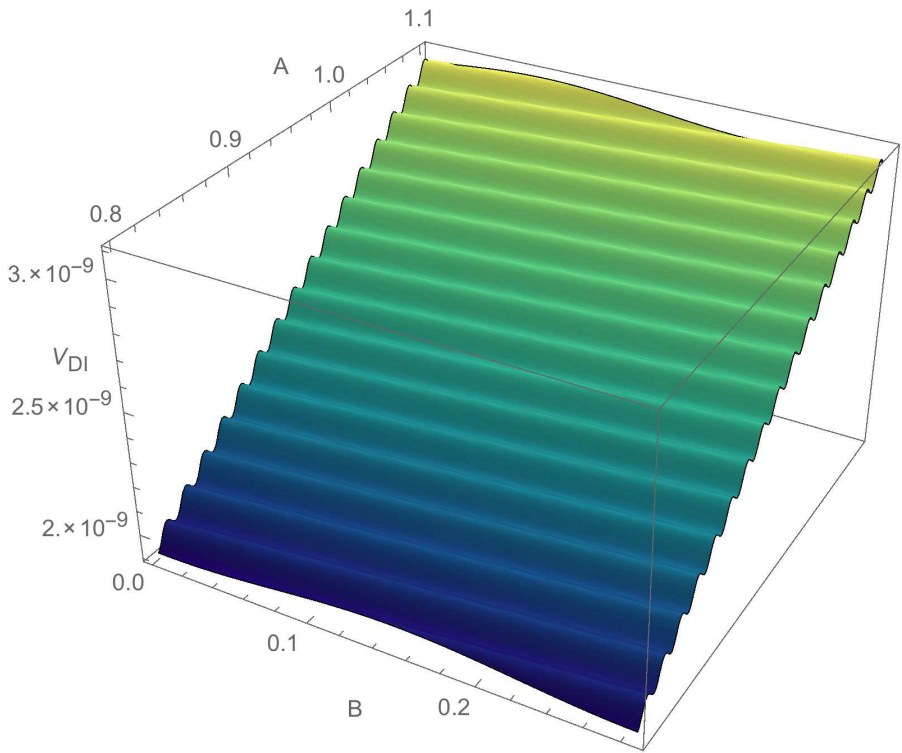}
  \vspace*{-0cm} 
\caption{The plot of $V_{DI}(A,B)$ for a typical values of 
parameters.
In the plot the two ends of the $B$-axis
which correspond to
$B=0$ and $B = 2\pi f_B$ are identified.}
 \vspace{0cm}
 \label{fig:DIpot}
\end{figure}

By taking the leading $n=1$ term in (\ref{Vf}), 
we obtain the potential for Dante's Inferno model
(\ref{VDI}):
\begin{equation}
V_{DI}(A,B) 
= 
V_A(A) 
+ 
\Lambda^4 
\left\{
1 - \cos 
 \left(
 \frac{A}{f_A} - \frac{B}{f_B}
 \right)
\right\},
\label{VDI2}
\end{equation}
with
\begin{equation}
V_A(A) = \frac{m^2}{2} A^2 - \frac{\lambda}{4!} A^4 ,
\label{VA2}
\end{equation}
and
\begin{equation}
\Lambda^4 = \frac{3}{\pi^2 (2 \pi L_5)^4} .
\label{Lambda4}
\end{equation}
The plot of the potential with typical values of 
parameters is shown in Fig.~\ref{fig:DIpot}.

To describe inflation in
Dante's Inferno model,
it is convenient to make a rotation in the field space
\cite{Berg:2009tg}:
\begin{equation}
\left(
\begin{array}{c}
	\tilde{A} \\
	\tilde{B}
\end{array}
\right)
=
\left(
\begin{array}{cc}
\cos \gamma & - \sin \gamma\\
\sin \gamma & \cos \gamma
\end{array}
\right)
\left(
\begin{array}{c}
  A \\
	B
\end{array}
\right) ,
\label{rot}
\end{equation}
where
\begin{equation}
\sin \gamma = \frac{f_A}{\sqrt{f_A^2 + f_B^2}},
\quad
\cos \gamma = \frac{f_B}{\sqrt{f_A^2 + f_B^2}}  .
\label{angle}
\end{equation}
In terms of the rotated fields, 
the potential (\ref{VDI2})
becomes
\begin{equation}
V_{DI}(\tilde{A},\tilde{B})
=
\frac{m^2}{2} (\tilde{A} \cos \gamma + \tilde{B} \sin \gamma)^2 
- \frac{\lambda}{4!} (\tilde{A} \cos \gamma + \tilde{B} \sin \gamma)^4 
+ 
\Lambda^4 
\left(
1 - \cos \frac{\tilde{A}}{f}
\right),
\label{tildepot}
\end{equation}
where
\begin{equation}
f = \frac{f_A f_B}{\sqrt{f_A^2 + f_B^2}} .
\label{f}
\end{equation}
Now, the following two conditions 
are imposed
in Dante's Inferno model:
\begin{align}
\mbox{condition 1}& \quad f_A \ll f_B \lesssim 1.
\label{cond1}\\
\mbox{condition 2}& \quad 
\left|
\pa_{\tilde{A}} V_A (A) |_{A=A_{in}} 
\right|
\ll \frac{\Lambda^4}{f}  .
\label{cond2}
\end{align}
Here, $A_{in}$ is the value of the field $A$ 
when the inflation started.
The first hierarchy in the 
condition~1 
is used to achieve effective super-Planckian
inflaton travel and
implies
\begin{equation}
\cos \gamma \simeq 1, \quad
\sin \gamma \simeq \frac{f_A}{f_B}, \quad
f \simeq f_A .
\label{conseq1}
\end{equation}
In terms of the variables 
in the five-dimensional gauge theory,
the condition~1 corresponds through (\ref{fAfB}) to
the hierarchy between the couplings of the different gauge groups
\cite{Furuuchi:2014cwa}:
\begin{equation}
g_B \ll g_A .
\label{gBgA}
\end{equation} 
The second inequality
in the condition~1 
is motivated
by the weak gravity conjecture \cite{ArkaniHamed:2006dz},
as mentioned in the introduction.
From (\ref{fAfB}) this condition amounts to
\begin{equation}
2\pi L_5 \gtrsim \frac{1}{g_B} .
\label{L5gBMP}
\end{equation}
The condition~2 (\ref{cond2})
is 
a requirement
for the field $\tilde{A}$ to
roll down to $\tilde{B}$-dependent local minimum 
much faster than the field $\tilde{B}$,
which is to be identified with the inflaton,
rolls down.
It imposes the following condition
on the parameters of the five-dimensional gauge theory:
\begin{equation}
m^2 A_{in} - \frac{\lambda}{3!} A_{in}^3 
\ll 
\frac{\Lambda^4}{f_A}
= \frac{3 g_A}{\pi^2 (2\pi L_5)^3} .
\label{cond22}
\end{equation}
After $\tilde{A}$ settles down to
$\tilde{B}$ dependent local minimum,
the motion of $\tilde{B}$ leads to the 
slow-roll inflation.
By redefining $\tilde{B} = \phi$,
we obtain the following inflaton potential:
\begin{align}
V_{eff}(\phi) 
= V_A \left( \sin \gamma \tilde{B} \right)
&= \frac{m_{eff}^2}{2} \phi^2 - \frac{\lambda_{eff}}{4!} \phi^4 \nn\\
&= \frac{m_{eff}^2}{2} \phi^2 \left( 1 - c \phi^2 \right),
\label{Veff}
\end{align}
where
\begin{align}
m^2_{eff} &:= \sin^2\! \gamma \, m^2
\simeq \left( \frac{f_A}{f_B} \right)^2 m^2,
\label{meff}\\
\lambda_{eff} 
&:=
\sin^4\! \gamma \, \lambda
\simeq
\left( \frac{f_A}{f_B} \right)^4 \lambda ,
\label{lambdaeff}
\end{align}
and
\begin{equation}
\quad c := \frac{\lambda_{eff}}{12m_{eff}^2}.
\label{c}
\end{equation}
Compared with the original potential $V_A(A)$ of the field $A$,
the potential $V_{eff}(\phi)$ 
of the inflaton $\phi$ is stretched in the 
field space direction,
due to the rotation in the field space (\ref{rot}),
see Fig.~\ref{fig:DIpot}.
This is the essential feature of the Dante's Inferno model
which allows the super-Planckian excursion of the inflaton
while the field ranges of the original fields $A$ and $B$ are sub-Planckian.

The inflaton potential
(\ref{Veff}) is not bounded from below,
but we will only consider the region of $\phi$ 
before the potential starts to go down:
\begin{equation}
|\phi | < |\phi|_{max} = \frac{1}{\sqrt{2c}}.
\label{phimax}
\end{equation}
We will not worry about
the potential beyond $|\phi|>|\phi|_{max}$
because this field region is not relevant for the inflation,
as we confirm shortly.
Actually, in \cite{Hashimoto:2008tw}
it has been shown that
massive vector field theories 
which can be embedded to a UV theory
whose $S$-matrix satisfies unitarity and
canonical analyticity constraints
do not have a
Lorentz-symmetry-breaking vacuum.
In such theories,
before the potential starts to go down,
the contribution 
from higher order terms in the potential
should come in to prevent Lorentz-symmetry-breaking local minimum,
assuming that the potential is bounded from below.

Since the inflaton potential (\ref{Veff})
is symmetric under the reflection $\phi \rightarrow -\phi$,
without loss of generality we assume $\phi \geq 0$ below.

\begin{figure}[t!]
 \centering
 \vspace{-0cm}
  \includegraphics[width=12cm]{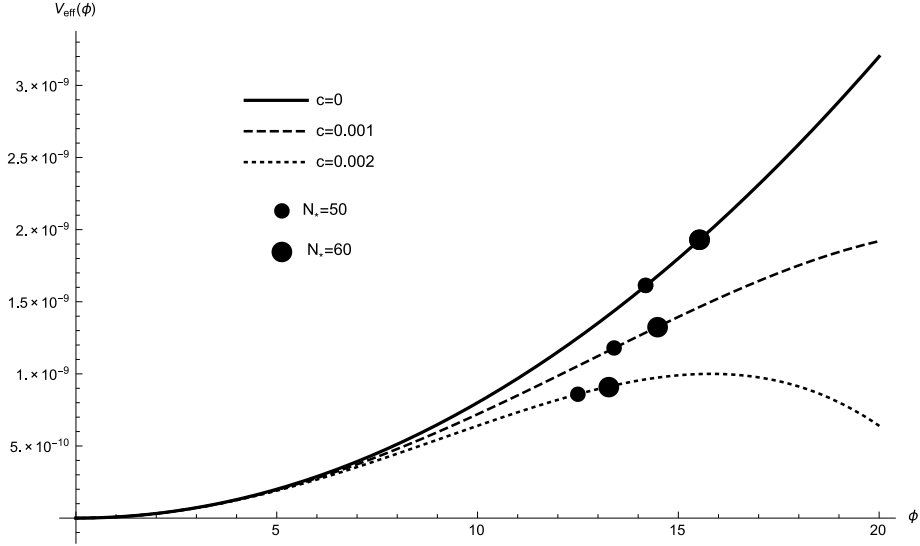}
  \vspace*{-0cm}
\caption{The effective potential 
(\ref{Veff}) and $\phi_\ast$
for $N_\ast=50$ and $N_\ast=60$
for different values of $c$.}
 \vspace{0cm}
 \label{fig:pot}
\end{figure}

We would like to compare 
our model with the CMB observations.
In order for that, 
we should impose the following condition:
\begin{equation}
\mbox{condition 3}\qquad \pa_{\tilde{A}}^2 V_{DI}(\tilde{A},\tilde{B}) 
\gg H^2,
\label{cond3}
\end{equation}
during the inflation, where
$H = \dot{a}(t)/a(t)$, with the dot denoting 
the derivative
with respect to the time $t$.
If this condition is satisfied,
and there is no other light scalar 
field with mass below $H$ which we assume to be the case,
only inflaton contributes to the scalar power spectrum.
Taking into account the condition~2 (\ref{cond2}),
the condition~3 (\ref{cond3}) reduces to
\begin{equation}
\pa_{\tilde{A}}^2 V_{DI}(\tilde{A},\tilde{B})
\simeq
\frac{\Lambda^4}{f^2} 
\simeq
\frac{\Lambda^4}{f_A^2}
\simeq
\frac{3 g_A^2}{\pi^2(2\pi L_5)^2}
\gg H^2 .
\label{cond32}
\end{equation}
This condition will be examined further later.

From the inflaton potential (\ref{Veff}),
the slow-roll parameters are calculated as
\begin{align}
\epsilon(\phi)
&:=
\frac{1}{2}\left(\frac{V'_{eff}}{V_{eff}}\right)^2
=
\frac{2}{\phi^2}
\left(\frac{1-2c\phi^2}{1-c\phi^2}
\right)^2 ,
\label{ep}\\
\eta(\phi)
&:=
\frac{V''_{eff}}{V_{eff}}
= \frac{2}{\phi^2}\frac{1 - 6 c \phi^2}{1-c \phi^2}.
\label{eta}
\end{align}
The spectral index is given as
\begin{equation}
n_s = 1 - 6 \epsilon(\phi_\ast) + 2 \eta(\phi_\ast),
\end{equation}
where
the subscript $\ast$ refers to the value at
the pivot scale $0.002$ Mpc$^{-1}$, for
which we follow the Planck 2015 analysis \cite{Ade:2015xua}.
The number of e-fold is given as
\begin{align}
N(\phi)
&=
\int_{\phi_{end}}^{\phi} d\phi \, \frac{V_{eff}}{V'_{eff}}
= 
\int_{\phi_{end}}^{\phi} d\phi \,
\frac{\phi}{2}
\frac{1 - c \phi^2}{1-2c\phi^2}
\nn\\
&=
\left[
\frac{\phi^2}{8}
-
\frac{\ln (1-2 c \phi^2)}{16c}
\right]_{\phi_{end}}^{\phi} ,
\label{N}
\end{align}
where we have defined $\phi_{end}$
as the field value
when $\epsilon(\phi)$ 
first reaches $1$
after the inflation starts.
In the parameter region we will consider,
this will be determined dominantly by
the quadratic part of the potential and
given as
\begin{equation}
\phi_{end} \simeq \sqrt{2} .
\label{phiend}
\end{equation}
The scalar power spectrum is given by
\begin{align}
P_s
=
\frac{V_{eff}(\phi_\ast)}{24\pi^2 \epsilon(\phi_\ast)}
=
2.2 \times 10^{-9},
\label{Ps}
\end{align}
where the value in the right hand side is 
from the observation \cite{Ade:2015xua}.
The tensor-to-scalar ratio is given as	
\begin{equation}
r_\ast = 16 \epsilon (\phi_\ast) .
\label{r}
\end{equation}

\begin{figure}[t!]
 \centering
 \vspace{-0cm}
  \includegraphics[width=14cm]{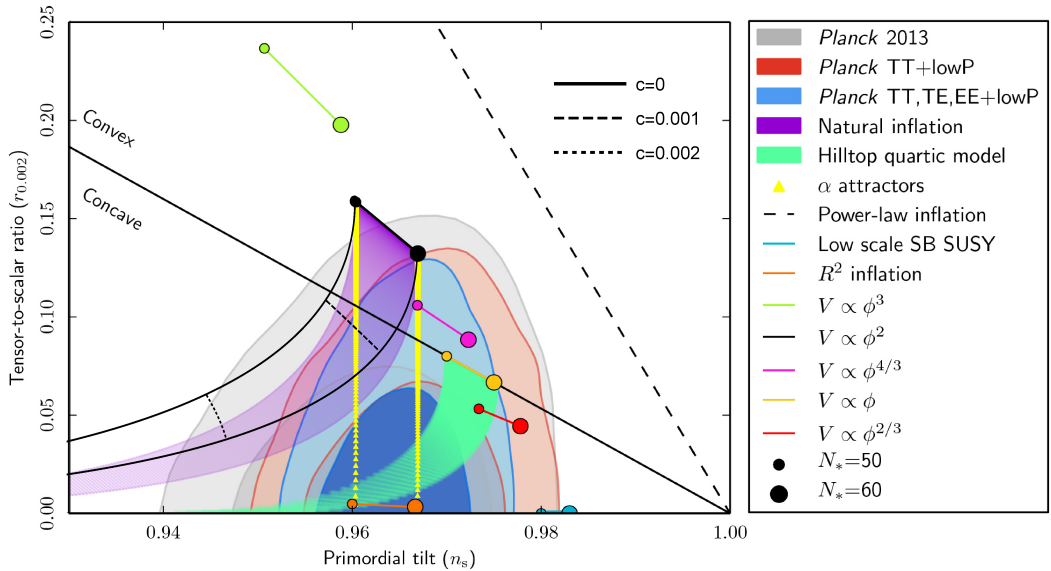}
  \vspace*{-0cm} 
\caption{Contour plots of $n_s-r$
for the inflation with potential (\ref{Veff}), 
with varying $c$ and with $N_\ast=50$ and $N_\ast=60$.
Compared with the Planck 2015 results \cite{Ade:2015lrj}.}
 \vspace{0cm}
 \label{fig:nsr}
\end{figure}

After obtaining the inflaton potential
(\ref{Veff}),
our model has two parameters 
$m_{eff}$ and $c$ in the potential (\ref{Veff}),
and one choice for the initial condition $\phi_\ast$.
The observational value of the 
power spectrum (\ref{Ps}) gives one relation among them,
and when the number of e-fold $N$ is specified,
(\ref{N}) gives another relation.
Then we are left with one independent parameter, 
for which we choose $c$.
The parameter $c$ is further constrained
by the observational bounds on 
the spectral index $n_s$ and
the tensor-to-scalar ratio $r$, as shown in 
the $n_s-r$ plane in Fig.~\ref{fig:nsr}
compared with that given 
in the Planck 2015 results \cite{Ade:2015lrj}.

From Fig.~\ref{fig:nsr}, we observe that
the inclusion of the quartic term in the potential
parametrized by positive $c$ of order $\Ord(10^{-3})$ 
pushes the model to the observationally favored direction.
This is quite as expected:
Positive $c$ results from
$v_2$ and $v_4$ both being negative (\ref{v2v4}).
Recall that (\ref{v2v4})
was a condition for avoiding the
IR obstruction to UV completion.
From this it follows that for larger $c$ 
the potential (\ref{Veff}) becomes lower at
large inflaton field values, 
as shown in Fig.~\ref{fig:pot}. 
This leads to smaller $r$ through (\ref{Ps}),
which is favored in the latest observations.

The main aim of Dante's Inferno model 
is to achieve super-Planckian inflaton excursion
in effective field theory while
the field ranges of the original fields
are sub-Planckian.
Thus we further require
\begin{equation}
\mbox{condition 4}\qquad
A_\ast \lesssim 1.
\label{cond4}
\end{equation}
From Fig.~\ref{fig:pot},
we observe that
$\phi_\ast
\simeq
12 \sim 15$
in the range of the parameter
$c$ of our interests.
Thus
\begin{equation}
A_{\ast} \simeq
\frac{f_A}{f_B}\phi_\ast
\gtrsim
\frac{g_B}{g_A}
\times 15 .
\label{Aast}
\end{equation}
Therefore, the condition~4 amounts to
\begin{equation}
g_A \gtrsim 15 g_B .
\label{gA15gB}
\end{equation}
This is compatible with the
condition~1, (\ref{cond1}).

\begin{figure}[t!]
 \centering
 \vspace{-0cm}
  \includegraphics[width=12cm]{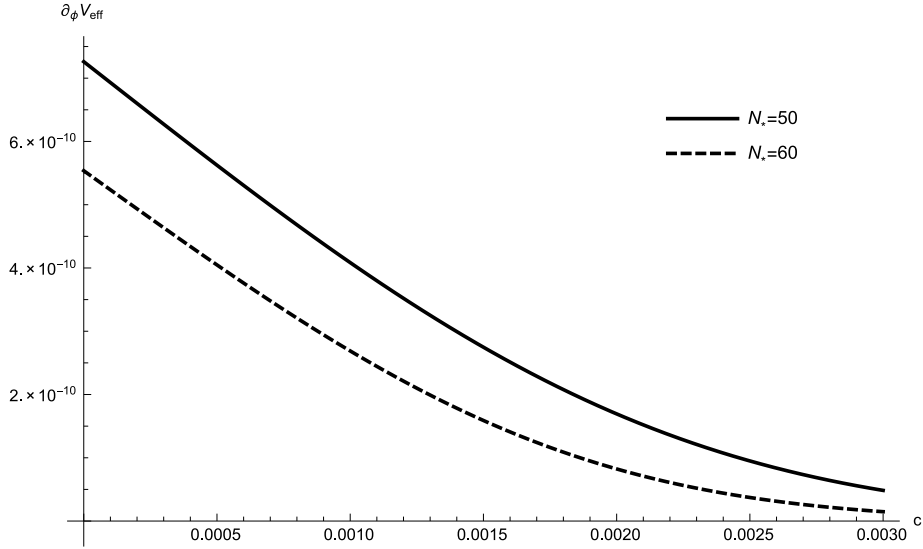}
  \vspace*{-0cm} 
\caption{The plot of $\partial_\phi V_{eff}(\phi_\ast)$
as a function of $c$.}
 \vspace{0cm}
 \label{fig:cdV}
\end{figure}

\begin{figure}[t!]
 \centering
 \vspace{-0cm}
  \includegraphics[width=12cm]{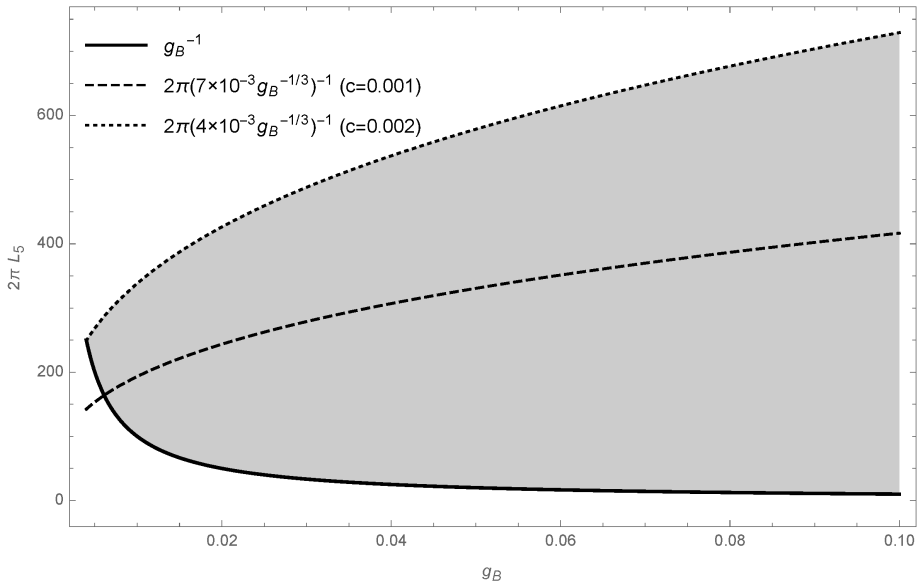}
  \vspace*{-0cm} 
\caption{The constraints 
on the compactification radius $L_5$ 
as a function of $g_B$.}
 \vspace{0cm}
 \label{fig:L}
\end{figure}

Next we would like to examine the condition~2.
Substituting (\ref{meff}) and (\ref{lambdaeff}) 
into (\ref{cond2}), 
we obtain
\begin{equation}
\frac{3g_B}{\pi^2(2\pi L_5)^3}
\gg 
m_{\rm eff}^2\phi_*
-\frac{\lambda_{\rm eff}}{6}\phi^3_*
=
\pa_\phi V_{eff}(\phi_\ast).
\label{gLdV}
\end{equation}
We have used $A_{in} \sim A_\ast$ in the above estimate.
As an example, we take $N_\ast=60$, $c=0.001$ case
which is observationally favorable as shown in
Fig.~\ref{fig:nsr}.
Then from Fig.~\ref{fig:cdV} we have
$\pa_\phi V_{eff} (\phi_\ast) \sim  4\times 10^{-10}$.
Putting this value into (\ref{gLdV}), we obtain
\begin{equation}
\frac{1}{L_5^3}\gg  3 \times 10^{-7} g_B^{-1},
\quad (N_\ast=60,\, c = 0.001),
\label{gBL53} 
\end{equation}
or equivalently
\begin{equation}
\frac{1}{L_5} > 7 \times 10^{-3} g_B^{-1/3}, 
\quad (N_\ast=60,\, c = 0.001).
\label{gB13L5}
\end{equation}
On the other hand, from the condition~1 (\ref{cond1})
we have 
\begin{equation}
2\pi L_5\gtrsim g_B^{-1}.
\label{L5gB}
\end{equation} 
Thus we arrive at 
\begin{equation}
7 \times 10^{-3}
g^{-1/3}_B
<
 \frac{1}{L_5}
\lesssim
2\pi g_B,
\quad (N_\ast=60,\, c = 0.001) .
\label{L5range}
\end{equation} 
Fig.~\ref{fig:L}
shows the allowed values of $L_5$ in (\ref{L5range}).
This figure should be looked together with
the condition (\ref{L5lb}), $2\pi L_5 \gtrsim 1 \times 10^2$,
which we have imposed to justify 
neglecting the contribution from the gauge field
$V_g(A)$ to the one-loop effective potential.
This condition still leaves a large portion
of the allowed parameter space.
Note that a natural value for $g_A$ is
$g_A \lesssim \Ord(1)$,
and through (\ref{gA15gB}) 
it means $g_B \lesssim \Ord(10^{-1})$.
As shown in Fig.~\ref{fig:L},
$L_5$ has allowed region in such values of
$g_B$.

\begin{figure}[t!]
 \centering
 \vspace{-0cm}
  \includegraphics[width=12cm]{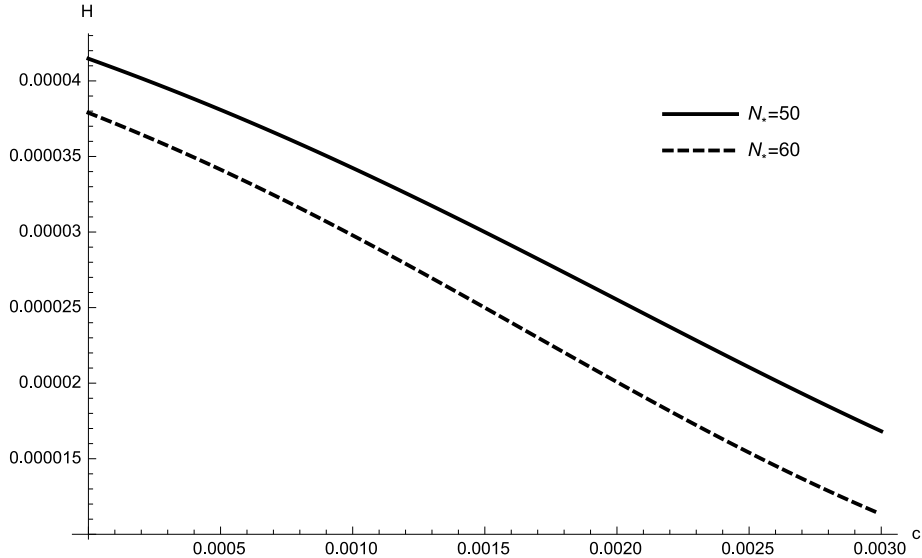}
  \vspace*{-0cm} 
\caption{The plot of $H_\ast$ as a function of $c$.}
 \vspace{0cm}
 \label{fig:cH}
\end{figure}

Finally, let us look back 
the condition~3 (\ref{cond3}).
From Fig.~\ref{fig:cH}, we observe
$H_\ast \sim \Ord(10^{-5})$.
Then (\ref{cond3}) gives only a very mild constraint 
$g_A \gg \Ord(10^{-5})$,
which is weaker than the bound given from
Fig.~\ref{fig:L} and (\ref{gA15gB}).

\section{Dante's Inferno model with
DBI action of a 5-brane}%
\label{secDBI}

In this section we study Dante's Inferno model
with a potential for the massive vector field
obtained from DBI action of a 5-brane.
Our main purpose in this section is to study 
the effects of higher order terms in the potential
comparing with the quartic potential in the previous section,
in an explicit model arising from string theory
whose $S$-matrix satisfies unitarity and
canonical analyticity constraints.
The low energy effective DBI action of a
D5-brane and that of an NS5-brane 
have been used in the axion monodromy inflation
\cite{Silverstein:2008sg,McAllister:2008hb},
and it is also interesting to 
observe the difference of the Dante's Inferno model with 5-branes.

In the case of D5-brane, the action is given as
\begin{equation}
S_{D5}
=
- T_{D5}
\int 
d^6\sigma
\sqrt{
- \det
\left( G_{ab} + {\cal F}_{ab} \right)
},
\quad
(a,b = 0,1,2,3,5,6),
\label{DBI}
\end{equation}
where
\begin{equation}
{\cal F}_{ab} = B_{ab} - \pa_a C_b + \pa_b C_a .
\label{calF}
\end{equation}
In (\ref{DBI}), $G_{ab} = G_{MN}\pa_a X^M \pa_b X^N$
and $B_{ab} = B_{MN} \pa_a X^M \pa_b X^N$
are the pull-back of the target space metric
and the NS-NS $2$-form field
to the D5-brane worldvolume, respectively.
$C_a$ is a $1$-form gauge field on the D5-brane.
The tension of the D5-brane is given by
\begin{equation}
T_{D5}
= \frac{1}{(2\pi)^5 g_s \alpha'^3} .
\label{TD5}
\end{equation}
The NS-NS $2$-form field $B_{MN}$ also has the kinetic term
in six-dimension which follows from a compactification
of the kinetic term in ten-dimensional bulk:
\begin{eqnarray}
\frac{1}{2(2\pi)^7 g_s^2 (\alpha')^4}
\int d^{10}x\,
H_{MNL}H^{MNL},
\label{Bkin}
\end{eqnarray}
\begin{eqnarray}
H_{MNL} = \partial_{[M} B_{NL]},
\label{H}
\end{eqnarray}
where $[\ldots]$ denotes the antisymmetrization.
The normalization of the kinetic term in six-dimension
depends on the volume of the compactified four-dimensional space.
Since it is simpler to directly 
discuss the normalization of the four-dimensional kinetic term
after further compactification of two more directions
as will be done below,
we don't explicitly write down the 
six-dimensional kinetic term here.

Let us consider the background
\begin{equation}
G_{MN} = \mbox{diag} (+-----),
\quad
B_{MN} = 0 ,
\label{BG}
\end{equation}
in the static gauge
$\sigma^a = x^a$ $(a = 0,1,2,3,5,6)$.\footnote{%
We will turn on the zero-mode of $B_{56}$ later.}
In the perturbative expansions in string coupling,
the constant shift of the NS-NS $2$-form field
\begin{equation}
B_{56} \rightarrow 
B_{56} + 2\pi \frac{2\pi \alpha'}{(2\pi L_5)(2\pi L_6)}  ,
\label{Bshift}
\end{equation}
is a symmetry.
The shift symmetry (\ref{Bshift}) is 
broken in the existence of the D5-brane.
(If we consider all the winding sectors,
the shift (\ref{Bshift}) exhibits monodromy.)
Upon double dimensional reduction
along the sixth direction,
the zero-mode of $B_{M6}$ becomes
a massive vector field in five-dimension
which we denote as $a_M$,
whereas
the zero-mode of the gauge field $C_M$
becomes the Stueckelberg scalar field 
which we denote as $\Theta$:
\begin{equation}
{\cal F}_{a6} = 
\frac{2\pi \alpha'}{(2\pi L_5)(2\pi L_6)}
\left(
a_M
- 
\pa_M \th
\right)
=
\frac{2\pi \alpha'}{(2\pi L_5)(2\pi L_6)}
\ca_M ,
\label{BM6}
\end{equation}
where
we will identify
$a_M$ and $\ca_M$
with
$A_M$ and $\cA_M$ in the previous section
up to a proportionality constant,
respectively.
We will fix the proportionality constants shortly.
The five-dimensional potential for $\ca_M$ 
after the double dimensional reduction is given by
\begin{equation}
\rho
T_{D5} (2\pi L_6)
\int d^5x
\sqrt{1 - \frac{(2\pi \alpha')^2}{(2\pi L_5)^2(2 \pi L_6)^2} \ca_M \ca^M} ,
\quad (M,N = 0,1,2,3,5).
\label{5D}
\end{equation}
Here, $\rho$ represents the numerical factor which
depends on the detail of the 
six-dimensional compact space,
possibly with a warp factor.

Now we would like to have 
a closer look at the IR obstruction to UV completion
for the massive vector field theory of $\cA_M$.
In \cite{Adams:2006sv},
DBI action was taken as
an example which is free from
the IR obstruction.                   
\cite{Adams:2006sv} focused on
the embedding coordinate fields $X^I(x)$.
When there is a small dimensionless expansion parameter, 
$g_s$ in this case,
unitarity and analyticity of the forward scattering
constrains
the sign of the coefficients of $(\pa_N X^I \pa^N X^I)^n$
to be all positive in the action \cite{Adams:2006sv}.
A prescription suggested in \cite{Hashimoto:2008tw}
for massive vector field models
was that the constraints from the
IR obstruction to UV completion 
on the sign of the coefficient of 
the term $(\cA_N \cA^N)^n$ is identical to
that on the coefficient of the term $(\pa_N X^I \pa^N X^I)^n$.
The five-dimensional action obtained
from the six-dimensional DBI action satisfy these conditions,
as can be checked from the Taylor expansion of 
the potential (\ref{5D}).
As DBI action is a low energy effective action
derived from string theory whose $S$-matrices
satisfy unitarity and canonical analyticity constraints, 
this supports the prescription
proposed in \cite{Hashimoto:2008tw}.

By further double dimensional reduction along the fifth direction, 
we obtain the four-dimensional potential for the field $a$
which is the zero-mode of $\ca_4$:
\begin{equation}
V_a (a)
=
\rho T_{D5} (2\pi L_5)(2\pi L_6)
\int d^4x
\sqrt{
1 +
\frac{(2\pi \alpha')^2}{(2\pi L_5)^2(2\pi L_6)^2} a^2
}.
\label{DBIVa}
\end{equation}
Recall our metric convention (\ref{metric}).
Here,
$a$ is normalized so that
$a \rightarrow a + (2\pi)$ corresponds to
the shift symmetry (\ref{Bshift}).
Thus if we identify this potential 
with $V_A(A)$ of the Dante's Inferno potential (\ref{VDI}),
the proportionality constant 
between the field $A$ and $a$ is fixed as
\begin{equation}
A = {f_A} a .
\label{Aa}
\end{equation}
This means that the kinetic term of $a$ was given as
\begin{equation}
\int d^4x\, \frac{f_A^2}{2} \pa_\mu a \pa^\mu a .
\label{akin}
\end{equation}
The four-dimensional kinetic term (\ref{akin}) follows from
the compactification of the 
ten-dimensional kinetic term (\ref{Bkin}),
and $f_A$ depends on the volume of the compactified space.

In terms of the field $A$,
the potential (\ref{DBIVa}) is written as
\begin{equation}
V_A (A)
=
\rho
T_{D5} (2\pi L_5)(2\pi L_6)
\int d^4x
\sqrt{
1 +
\frac{(2\pi \alpha')^2}{(2\pi L_5)^2(2\pi L_6)^2 f_A^2} A^2
}.
\label{DBIVA}
\end{equation}
From (\ref{DBIVA})
we read off the convergence radius $A_c$
for the Taylor expansion around $A=0$: 
\begin{equation}
A_c = \frac{f_A (2\pi L_5)(2\pi L_6)}{2\pi \alpha'}.
\label{Ac}
\end{equation}
As we should assume that 
$(2\pi L_5), (2\pi L_6) \gg (\alpha')^{1/2}$
in order to justify the 
suppression of the string corrections,
we have
\begin{equation}
A_c \gg \frac{f_A}{2\pi}.
\label{AcfA}
\end{equation}

Now, we would like to study Dante's Inferno model
as we have done in the previous section for
the classical potential (\ref{DBIVA}).
In order for that, we assume that there is also
the field $B$ in (\ref{VDI}) which may arise from 
form fields in higher dimensions,\footnote{For example,
NS-NS two-form field $B_{MN}$ with $M=5$ and $N$ in
other extra dimensional direction may do the job.} 
and that the sinusoidal potential in (\ref{VDI}) is also generated
in the same way as in the previous section.
The essential ingredient of
Dante's Inferno model is that
the effective potential 
for the inflaton field $\phi = \tilde{B}$
is stretched 
by the factor $1/\sin \gamma \simeq f_B/f_A$
in the field space direction
compared with the potential for the field $A$:
\begin{align}
V_{eff}(\phi) 
&=
V_A (\sin \gamma \phi)
\simeq
V_A \left(\frac{f_A}{f_B} \phi  \right) 
\nn\\
&=
\rho
T_{D5} (2\pi L_5)(2\pi L_6)
\int d^4x
\sqrt{
1 +
\frac{(2\pi)^2\alpha'^2}{(2\pi L_5)^2(2\pi L_6)^2 f_B^2} \phi^2
}
\nn\\
&=
\rho
T_{D5} (2\pi L_5)(2\pi L_6)
\int d^4x
\sqrt{
1 +
\left(
\frac{\phi}{\phi_c}
\right)^2
},
\label{Vstretched} 
\end{align}
where the convergence radius $\phi_c$
for the Taylor expansion
is given by
\begin{equation}
\phi_c 
= \frac{f_B}{f_A} A_c
=
\frac{f_B (2\pi L_5)(2\pi L_6)}{2\pi \alpha'} .
\label{phic}
\end{equation}
When $\phi_\ast \ll \phi_c$,
the Taylor expansion of the square root
is a good approximation
for describing the inflation, 
while when $\phi_\ast \gg \phi_c$
the potential (\ref{Vstretched})
is approximately a linear potential.
The latter was the case studied in
\cite{McAllister:2008hb}
for a single axion monodromy model.
We examine below which is the case
for the current model.

\begin{figure}[t!]
 \centering
 \vspace*{0cm}
 \hspace*{0cm}
  \includegraphics[width=12cm]{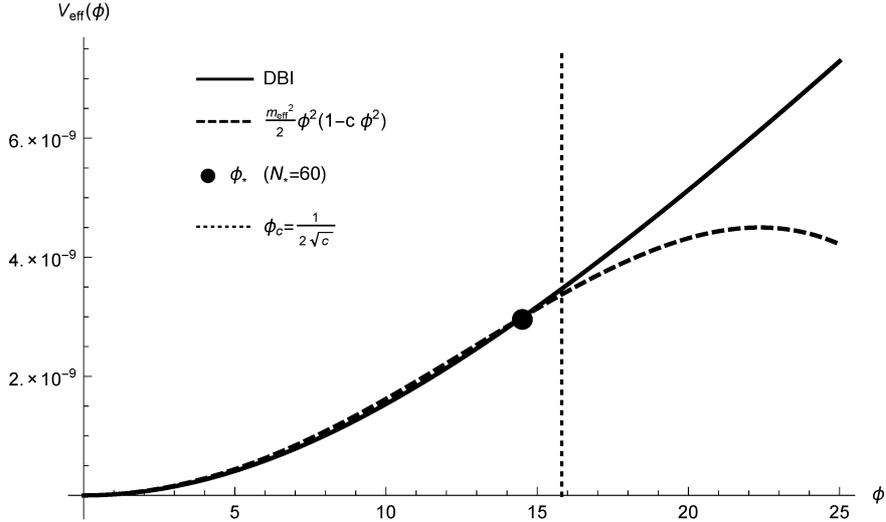}
  \vspace*{1.2cm} 
\caption{A comparison between the quartic potential
(\ref{Veff}) and the potential  
(\ref{Vstretched}) obtained from the DBI action.}
 \vspace{0cm}
 \label{fig:comparisonDBI}
\end{figure}

Let us first 
truncate the potential (\ref{Vstretched})
at the quartic order in the Taylor expansion
and apply the results in the previous section.
Then, the effective mass $m_{eff}$ and 
the effective quartic coupling constant $\lambda_{eff}$
in the truncated potential are given by
\begin{align}
m_{eff}^2 &= 
\rho
T_{D5}(2\pi L_5)(2\pi L_6)
\frac{1}{\phi_c^2}
\nn\\
&= 
\frac{\rho}{(2\pi)^2 g_s} \frac{(2\pi L_5)(2\pi L_6)}{(2\pi \alpha')^3}
\frac{1}{\phi_c^2} ,
\label{DBIm}
\end{align}
\begin{align}
\lambda_{eff}
&=
\frac{4!}{8}
\rho
T_{D5}(2\pi L_5)(2\pi L_6)
\frac{1}{\phi_c^4}
\nn\\
&=
\frac{3 \rho}{(2\pi)^2 g_s}
\frac{(2\pi L_5)(2\pi L_6)}{(2\pi \alpha')^3}
\frac{1}{\phi_c^4} .
\label{DBIlambda}
\end{align}
Thus we obtain
\begin{equation}
c 
= \frac{\lambda_{eff}}{12 m_{eff}^2} 
=\frac{1}{4 \phi_c^2}.
\label{DBIc}
\end{equation}
(\ref{DBIc}) gives 
\begin{equation}
\phi_c = \frac{1}{2\sqrt{c}} 
\lesssim |\phi|_{max} = \frac{1}{\sqrt{2c}},
\label{phicphimaxNS}
\end{equation}
where $|\phi|_{max}$ was given in (\ref{phimax}).
For example, for $c = 0.001$,
$\phi_c \simeq 16 
\gtrsim \phi_\ast \simeq 15$.
Therefore,
the inflation takes place within the convergence radius of
the Taylor expansion
in the truncated potential.
In Fig.~\ref{fig:comparisonDBI}
we compare the quartic potential (\ref{Veff})
and the potential following from the DBI action (\ref{Vstretched}).
In Fig.~\ref{fig:comparisonDBI}
the parameters of the potential (\ref{Vstretched}) are tuned
so that it coincides with the quartic potential (\ref{Veff})
at $\phi=0$ and $\phi=\phi_\ast$.
What this means is that
we regard the quartic potential (\ref{Veff})
not as a truncation of the
Taylor expansion of the potential (\ref{Vstretched})
at the quartic order,
but as a phenomenological parametrization to fit it.
We observe that this phenomenological parametrization
is a rather good approximation to the potential (\ref{Vstretched})
in the field range where the inflation takes place.
Thus although the inflation starts 
close to the convergence radius and
the quartic potential
may not be a very accurate approximation
to the potential (\ref{Vstretched}),
it will be more than enough for qualitative estimates.
It is interesting
that the inflation does not takes place in the field range
where the linear approximation at the large field value is valid,
which was the case in the 
well known examples of axion monodromy model
\cite{McAllister:2008hb}.
The difference originates from 
the very essence of Dante's Inferno model
that the potential
for the effective inflaton (\ref{Vstretched})
is stretched from that for the original field $A$.

Note that the potential obtained from DBI action
monotonically increases,
differing from the quartic potential which starts
to go down from $|\phi| = |\phi|_{max}$.
We do not expect the quartic potential to be
a good description in these regions,
which however are irrelevant when discussing inflation.

\begin{figure}[t!]
 \centering
 \vspace{-0cm}
  \includegraphics[width=12cm]{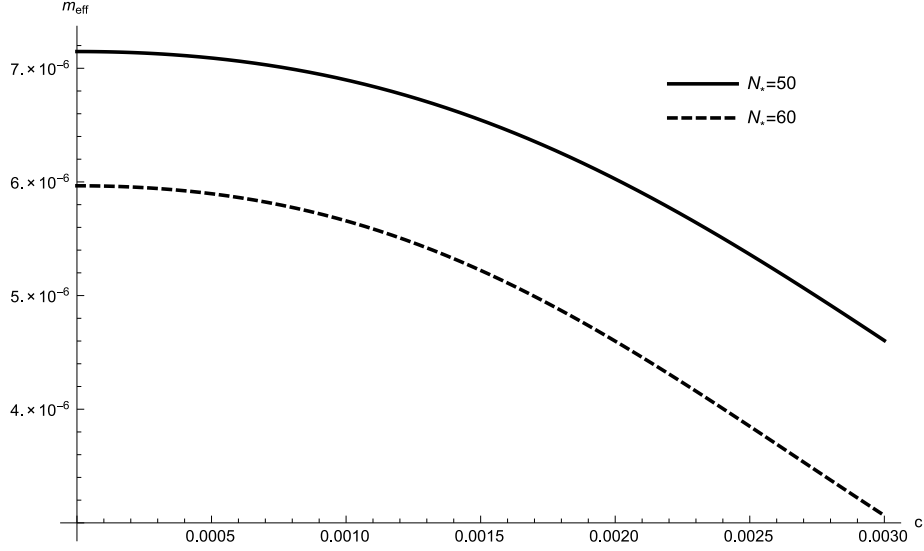}
  \vspace*{-0cm} 
\caption{The plot of $m_{eff}$
as a function of $c$.}
 \vspace{0cm}
 \label{fig:cm}
\end{figure}

\begin{figure}[t!]
 \centering
 \vspace{-0cm}
  \includegraphics[width=12cm]{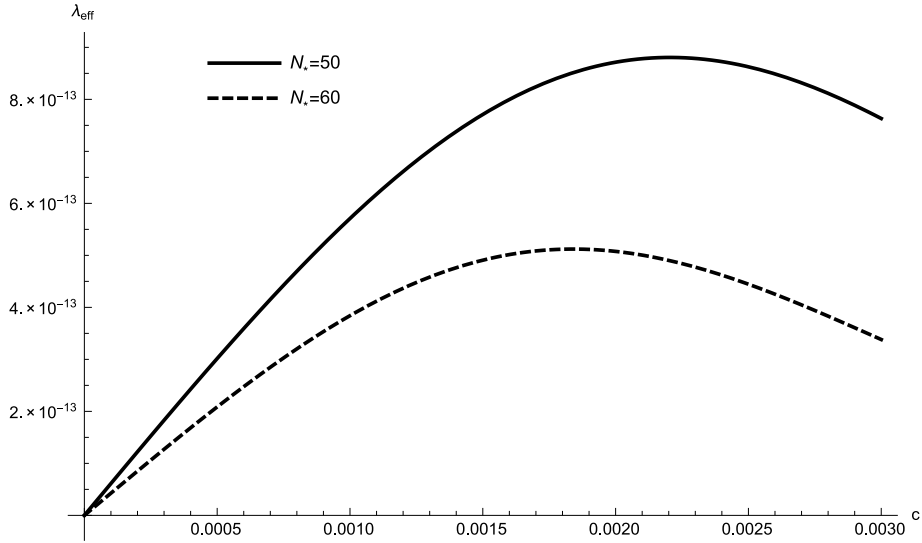}
  \vspace*{-0cm} 
\caption{The plot of $\lambda_{eff}$
as a function of $c$.}
 \vspace{0cm}
 \label{fig:clambda}
\end{figure}

As we have seen, though the truncation of the potential
at the quartic order in Taylor expansion
may not be a very precise description,
it should be good enough for qualitative estimates,
so let us proceed with the values obtained in the previous section.
From (\ref{phic}), 
the value $\phi_c \simeq 16$ can be achieved, for example,
$(2\pi L_5)^{-1} \lesssim (2\pi L_6)^{-1} \sim \Ord(10^{-2})$,
$(2\pi \alpha')^{-1} \sim \Ord (10^{-1})$ with $f_B \lesssim 1$.
For these values, we obtain
\begin{align}
m_{eff}^2 &\gtrsim 
\frac{\rho}{g_s} \times 10^{-1},
\label{DBImval}\\
\lambda_{eff}
&\gtrsim
\frac{3\rho}{g_s} \times 10^{-3}.
\label{DBIlambdaval}
\end{align}
From
Fig.~\ref{fig:cm} and
Fig.~\ref{fig:clambda},
$m_{eff}^2 \sim \Ord(10^{-11})$ and
$\lambda_{eff} \sim \Ord(10^{-13})$
for $c=0.001$.
Therefore,
${\rho}/{g_s} \lesssim \Ord(10^{-10})$
would realize successful Dante's Inferno model
from higher-dimensional gauge theory
discussed in the previous section.
This value of $\rho/g_s$ may be realizable
in appropriate warped geometries,
though the study of 
consistent realization in string theory
is beyond the scope of the current article.

The case of DBI action of
NS5-brane with RR 2-form field is similar, except
for the string coupling dependence.
Therefore, we just write down the corresponding formulas.
Instead of 
(\ref{TD5}), (\ref{phic}), (\ref{DBIm}), 
(\ref{DBIlambda}) and (\ref{DBIc}),
we have
\begin{equation}
T_{NS5}
=
\frac{1}{(2\pi)^5 g_s^2 \alpha'^3},
\label{TNS5}
\end{equation}
\begin{equation}
\phi_c =
\frac{f_B (2\pi L_5)(2\pi L_6)}{g_s (2\pi \alpha')} ,
\label{phicNS}
\end{equation}
\begin{align}
m_{eff}^2 &= 
\rho
T_{NS5}(2\pi L_5)(2\pi L_6)
\frac{g_s^2 (2\pi)^2\alpha'^2}{(2\pi L_5)^2(2\pi L_6)^2 f_B^2}
\nn\\
&= 
\frac{\rho}{(2\pi)^2}
\frac{1}{2\pi \alpha'}
\frac{1}{(2\pi L_5)(2\pi L_6)}
\frac{1}{f_B^2},
\label{DBImNS}
\end{align}
\begin{align}
\lambda_{eff}
&=
\frac{4!}{8}
\rho
T_{NS5}(2\pi L_5)(2\pi L_6)
\left(
\frac{g_s^2 (2\pi \alpha')^2}{(2\pi L_5)^2(2\pi L_6)^2 f_B^2}
\right)^2 
\nn\\
&=
\frac{3 \rho g_s^2}{(2\pi)^2}
\frac{2\pi \alpha'}{(2\pi L_5)^3 (2\pi L_6)^3 f_B^4} ,
\label{DBIlambdaNS}
\end{align}
and thus
\begin{equation}
c 
= \frac{\lambda_{eff}}{12 m_{eff}^2} 
=
\frac{1}{4 \phi_c^2},
\label{DBIcNS}
\end{equation}
or
\begin{equation}
\phi_c = \frac{1}{2\sqrt{c}} 
\lesssim |\phi|_{max} = \frac{1}{\sqrt{2c}}.
\label{phicphimax}
\end{equation}

\section{Summary and discussions}\label{secsd}

In this article, we spotted light 
on the constraints from the
IR obstruction to UV completion \cite{Adams:2006sv,Hashimoto:2008tw}
on inflation models obtained from higher-dimensional 
massive vector field theories.
While weak gravity conjecture \cite{ArkaniHamed:2006dz}
which also discusses constraints from UV completion
has been discussed extensively in the past few years,
the IR obstruction to UV completion in this context 
has not been studied previously.
In Dante's Inferno model, which is most promising in this class of models, 
we have shown that the constraint on the sign of the 
quartic term in the potential of the massive vector field
is in favor of 
the current observational upper bound on tensor-to-scalar ratio.
We also discussed DBI actions on 5-branes in string theory.
In particular, we studied the effects of higher order terms in the potential
of the massive vector field, 
which arises from 
NS-NS/RR two-form field on a 5-brane upon dimensional reduction.
Interestingly, in these models inflation takes place
within the convergence radius of the Taylor expansion.
This is in contrast to 
the well known examples of axion monodromy inflation \cite{McAllister:2008hb},
in which inflaton takes place outside the convergence radius
of the Taylor expansion.
The difference arises 
from the very essence of Dante's Inferno model
that the effective inflaton potential is
stretched in the inflaton field direction
compared with the potential for the original field.
The result also tells us
that the model with a potential up to the quartic order
can approximate the models with higher order terms
reasonably well.

While in \cite{Adams:2006sv} and \cite{Hashimoto:2008tw},
canonical analyticity constraints
on $S$-matrices in flat space-time were considered,
it has been pointed out that
analyticity structure of $S$-matrices
could be much richer in curved space-time
\cite{Hollowood:2007kt,Hollowood:2007ku,Hollowood:2008kq}.
It will be interesting to explore whether further
constraints on effective field theories arise
in curved space-time,
and if they do, what are the implications to inflation models,
in particular those
which are obtained from higher-dimensional 
massive vector field theories.\\[.1cm]

\begin{center}
\textbf{Acknowledgments}
\end{center}
Y.K. wishes to thank 
Carlos~Cardona, 
Chong-Sun~Chu, 
Dimitrios~Giataganas and Jackson~M.~S.~Wu for useful discussions. 
The work of Y.K. is supported in part by the National Center for Theoretical Sciences (NCTS) 
and the grant 101-2112-M-007-021-MY3 
of the Ministry of Science and Technology of Taiwan R.O.C.. 

\appendix

\section{The One-loop effective potential}\label{apponeloop}

\subsection{Calculation of the one-loop effective potential}

In this appendix,
we evaluate the 
one-loop effective potential 
for the zero-modes of $A_5$ and $B_5$.
We consider the action in five dimensions consists of 
the $U(1)$ gauge fields $A_M$ and $B_M$, 
a massless fermion $\psi$ with 
charges $\ell$ and $-\ell'$ of $U(1)_A$ and $U(1)_B$,
respectively, and the Stueckelberg field $\theta$
associated with the $U(1)_A$ gauge group:
\begin{align}
S
=
\int d^5x
&
\left[-\frac{1}{4} F^{(A)}_{MN} F^{(A)\, MN}
-\frac{1}{4} F^{(B)}_{MN} F^{(B)\, MN}
-V_{cl}(\cA_M)+{\bar \psi}i\Gamma^{M}D_M\psi \right],\nn\\
&\qquad \qquad \qquad \qquad (M,N = 0,1,2,3,5),
\label{appS}
\end{align}
where
\begin{equation}
F^{(A)}_{MN}=\partial_M A_N-\partial_N A_M,
\quad 
F^{(B)}_{MN}=\partial_M B_N-\partial_N B_M,
\end{equation}
\begin{equation}
\cA_M = A_M- g_{A5}\partial_M \theta,
\end{equation}
and
\begin{equation}
D_M\psi=\partial_M\psi-ig_{A5}\ell A_M\psi-ig_{B5}(-\ell')B_M\psi.
\end{equation}
We consider the following classical potential 
for $\cA_M$:
\begin{equation}
V_{cl}(\cA_M)=v_2\cA_M\cA^M+v_4(\cA_M\cA^M)^2.
\label{spo}
\end{equation}
The action (\ref{appS}) 
is invariant under the following gauge transformations:
\begin{align}
&A_M \rightarrow A_M + \partial_M \Lambda^{(A)},
\quad 
B_M \rightarrow B_M + \partial_M \Lambda^{(B)},\notag \\
&\psi \rightarrow e^{ i\ell g_{A5} \Lambda^{(A)} }
e^{ i(-\ell')g_{B5}\Lambda^{(B)} }\psi,
\quad 
\theta \rightarrow \theta + g^{-1}_{A5} \Lambda^{(A)}.
\end{align} 
We assume that the $U(1)_A$ gauge groups are compact, thus
the Stueckelberg field $\theta$ 
is periodically identified as
\begin{equation}
\theta \sim \theta+\frac{2\pi}{g_{A5}^2}.
\end{equation}
To evaluate the one-loop effective potential 
for the zero-modes of $A_5$ and $B_5$, 
we expand the action around $x$-independent classical values:
\begin{equation}
A_M(x)=A_M^c+A_M^q(x),\quad 
B_M(x)=B_M^c+B_M^q(x),\quad 
\psi=0+\psi^q(x),
\end{equation}
where $q$ denotes quantum fluctuation.
The expansion of the 
classical potential 
around the classical value of $\cA_M=\cA^c_M$ 
up 
to the quadratic order in the fluctuations
is given by
\begin{equation}
V_{cl}(\cA^c_M+\cA^q_M)=V_{cl}(\cA^c_M)
+
V_K(\cA^c_M)\cA^{qK}+\frac12 V_{KL}(\cA^c_M)\cA^{qK}\cA^{qL}+{\cal O}(\cA^{q3}_M),
\label{potex}
\end{equation}
where
\begin{align}
V_{cl}(\cA^c_M)&=v_2\cA^c_M\cA^{cM}+v_4(\cA^c_M\cA^{cM})^2,
\\
V_K(\cA^c_M)&=
\frac{\partial V_{cl}(\cA^c_M)}{\partial \cA^K}
=2v_2\cA^c_K+4v_4 \cA^c_M\cA^{cM}\cA^c_K
,\\
V_{KL}(\cA^c_M)&=
\frac{\partial^2 V_{cl}(\cA^c_M)}{\partial \cA^K\partial \cA^L}
=2v_2\eta_{KL}+4v_4(\eta_{KL}\cA^c_M\cA^{cM}+2\cA^c_K\cA^c_L).
\label{sede}
\end{align}
We also introduce the following gauge fixing term: 
\begin{equation}
S_{gf}=
\int d^5x
\left[
-\frac{1}{2\xi}(\partial_MA^{qM}+\xi m^2_{\cA}\theta^q)^2 
-\frac{1}{2\zeta} (\partial_MB^{qM})^2 
\right],
\label{rxi}
\end{equation} 
where 
\begin{equation}
m^2_{\cA} :=
-2v_2-4v_4 \cA^c_M\cA^{cM}.
\label{amass1}
\end{equation}
We shall choose $\xi=1$ and $\zeta=1$ in \eqref{rxi}.
Assuming ${\cA_M}$ is at the extremum of the potential, 
the action up to the quadratic order
in the fluctuations is given as
\begin{align}
S^{(2)}+S_{gf}
=
\int d^5x
&\left[
-\frac14 F^{(A)}_{MN}F^{(A) \, MN}
-\frac14F^{(B)}_{MN}F^{(B)\, MN}
+{\bar \psi}^qi\Gamma^{M}D_M(A^c_{M},B^c_{M})\psi^q\right.\notag\\
& 
-\frac12(\partial_MA^{qM}+g_{A5}m^2_{\cA}\theta^q)^2 
-\frac12 (\partial_MB^{qM})^2 \notag\\
&\left.
-\frac{1}{2}V_{KL}(\cA^c_M)(A^{qK}-g_{A5}\partial^K\theta^q)
(A^{qL}-g_{A5}\partial^L\theta^q)\right]\notag\\
=
\int d^5x
&\left[
\frac12X_aM^{ab}X_b+\frac12B^q_N\partial_M\partial^MB^{qN}
+
{\bar \psi}^qi\Gamma^{M}D_M(A^c_{M},B^c_{M})\psi^q
\right],\nn\\
&
\label{rxifey}
\end{align}
where
\begin{align}
X_a&:= (A^q_M,g^{-1}_{A5}\theta^q),\qquad a=M,\theta,\notag\\
M^{ab}&:=
\left(\begin{array}{cc}
\eta^{MN}
\left(
\partial^2_{M} + m_\cA^2
\right)
-8v_4\cA^{cM}\cA^{cN} 
& 8v_4 g_{A5}^2\cA^{cK}\cA^{cM}\partial_{K}\\
-8v_4 g_{A5}^2\cA^{cK}\cA^{cN}\partial_{K}
& - g_{A5}^4 m^2_\cA 
\left(
\partial^2_{M}
+\frac{8v_4}{m^2_\cA}\cA^{c}_K\cA^{c}_L\partial^K\partial^L
+m_\cA^2
\right)
\end{array}\right).\notag\\
\end{align}
We also need to consider the ghost action associated with $U(1)_A$ gauge fixing since it couples to the Vacuum Expectation Value (VEV) 
of $\cA_M$ via $m_\cA^2$ (\ref{amass1}) 
hence contributes to the one-loop effective potential. 
The ghost action corresponding 
to the gauge fixing (\ref{rxi}) is given as
\begin{equation}
S_{c_A}
=\int d^5x
\left[
-{\bar c}_A\left(\partial_M\partial^M + m^2_\cA\right)c_A
\right].
\end{equation}
The ghosts for the $U(1)_B$ gauge group, 
$c_B$ and $\bar{c}_B$, are free as usual
and decouple from the rest of the calculations.

The fifth dimension is compactified on $S^1$ with radius $L_5$. 
The mode expansions of the fields in the fifth direction
are given as
\begin{align}
A_M(x,x^5)
&=
\frac{1}{\sqrt{2\pi L_5}}
\sum^{\infty}_{n=-\infty}A^{(n)}_M(x)e^{i\frac{ n }{L_5}x^5},
\quad {\rm same\ for} \ B_M,\ \psi,\ c_A,
\notag\\ 
\theta(x,x^5) 
&=  \frac{x^5}{g_{A5}^2 L_5} w 
+ \sum_{n=-\infty}^{\infty} \theta^{(n)}(x) e^{i\frac{ n }{L_5}x^5},
\end{align}
where $\theta$ can have integer winding number $w$,
but it can be set to zero by a gauge transformation
$A_5\rightarrow A_5+ k/(g_{A5}L_5)$, 
$\theta\rightarrow \theta+ kx^5/(g^2_{A5}L_5)$
($k$ is an integer). 
In what follows we will fix $w=0$.
We consider 
the following VEVs for $A_5^{(0)}$ and $B^{(0)}_5$:
\begin{equation}
\langle{ A^{(0)}_5 \rangle} = A, 
\qquad 
\langle{ B^{(0)}_5 \rangle} = B, 
\end{equation}
and the other fields have zero expectation values.
Corresponding to the VEV of $A_M$, the VEV of $\cA_M$ is
\begin{equation}
\cA^c_{\mu}=0,\quad 
\cA^c_5
=\frac{1}{\sqrt{2\pi L_5}}\langle{A^{(0)}_5\rangle}
.
\end{equation}
Now we have the quadratic actions for the gauge fields:
\begin{equation}
S^{(2)}_{A,\theta}=\int d^4x\sum_{n=-\infty}^{\infty}
\frac12{\tilde X}^{(n)}_a M_4^{ab}{\tilde X}^{(-n)}_b ,
\label{gauquad}
\end{equation}
where
\begin{equation}
{\tilde X}^{(n)}_a := (A^{(n)}_M,{\tilde \theta}^{(n)}),
\qquad
{\tilde \theta}^{(n)} := 
\left(g^2_{A5} m^2_\cA (2\pi L_5)\right)^{1/2}\theta^{(n)},
\end{equation}
\begin{equation}
M_4^{ab} :=
\left(\begin{array}{cc}
\eta^{MN}
\left(
\partial^2_{\mu}+(\frac{n}{L_5})^2+m_\cA^2
\right)
-8v_4(\cA^{c}_5)^2\delta_5^M\delta_5^N 
& \frac{8v_4}{m_\cA}\frac{n}{L_5}(\cA^{c}_5)^2\delta_5^M\\
-\frac{8v_4}{m_\cA}\frac{n}{L_5}(\cA^{c}_5)^2\delta_5^N
& 
-\left(
\partial^2_{\mu}+(\frac{n}{L_5})^2
(1+\frac{8v_4}{m_\cA^2}(\cA^{c}_5)^2)
+m_\cA^2\right) 
\end{array}\right),
\notag\\
\end{equation}
and that for the ghost fields $c_A$ and $\bar{c}_A$:
\begin{equation}
S_{c_A}=\int d^4x
\left[
-\sum_{n=-\infty}^{\infty}{\bar c}^{(n)}_A
\left(\partial^2_\mu+\left(\frac{n}{L_5}\right)^2
+ m^2_\cA\right)c^{(n)}_A
\right],
\label{ghquad}
\end{equation}
and that for the fermion:
\begin{equation}
S^{(2)}_{\psi}=
\int d^4x
\sum_{n=-\infty}^{\infty}{\bar \psi}^{(n)}\left(i\Gamma^{\mu}\partial_{\mu}
+\ell g_{A}\Gamma^5 A 
-\ell' g_{B}\Gamma^5 B 
-\Gamma^5\frac{n}{L_5}\right)
\psi^{(n)}.
\label{ferquad}
\end{equation}
In the above, $\mu$ denotes the directions
in the uncompactified four-dimensional space-time 
and runs from $0$ to $3$.
Note that $B_M$ bosons do not contribute 
to the one-loop effective potential 
because they do not couple to the background fields 
at the one-loop level.

We observe that the fermion contribution $V_f(A,B)$ to 
the one-loop effective potential $V(A,B)$ 
is the same as the previous study 
\cite{Furuuchi:2014cwa}:
\begin{align}
V_f(A,B)
&={\rm Tr}
\left[
{\rm ln} \left(-i \Gamma^{\mu}\partial_{\mu}
-\ell g_{4A}\Gamma^5\langle{A^{(0)}_5\rangle}
+\ell' g_{4B}\Gamma^5\langle{B^{(0)}_5\rangle}+\Gamma^5\frac{ n}{L_5}\right)
\right]
\notag\\
&=\frac12{\rm Tr}
\left[
{\mbox{1}\hspace{-0.25em}\mbox{l}}_{4 \times 4} \,
{\rm ln} 
\left\{ -\partial^2_{\mu}+
\left(\frac{n}{L_5}-
\frac{1}{2\pi L_5}
 \left(\frac{A}{f_A}-\frac{B}{f_B}\right)
\right)^2
\right\}
\right],
\end{align}
where
\begin{equation}
f_A=\frac{1}{(2\pi g_{A} \ell L_5)},\quad 
f_B=\frac{1}{(2\pi g_{B} \ell' L_5)}.
\end{equation} 
Employing the $\zeta$ function regularization, we obtain
\begin{align}
V_f(A,B)
&=\frac{3}{\pi^2 (2\pi L_5)^4}\sum_{n=1}^{\infty}\frac{1}{n^5}\cos \left[n\left(\frac{A}{f_A}-\frac{B}{f_B}\right)\right].
\label{appVf}
\end{align}
In (\ref{appVf}) we have subtracted the constant part by hand.
Although the constant term has a physical significance,
the huge discrepancy between the theoretically natural value 
of the constant term
and the observationally suggested value of it
is
the notorious cosmological constant problem, 
which we do not attempt to address in this article.

Next we turn to the gauge boson contributions to the 
one-loop effective potential. 
We introduce the Euclidean time $\tau$ and the Euclidean gauge field $A_E^M$ as follows:
\begin{equation}
\tau=it,\quad A_E^0=iA^0,\quad A_E^i=A^i,\quad A_E^5=A^5  .
\end{equation}
The gauge boson loops give rise to the effective action of $A$:
\begin{equation}
\Gamma_g(A)=
-2\ln\left.\det D^2\right|_{A^E_{\mu}}
-\frac12\ln\left.\det D^2\right|_{A^E_5}
-\frac12\ln\left.\det D^2\right|_{\tilde \theta}+\ln\left.\det D^2 \right|_{c_A},
\end{equation}
where the determinant is that with respect to $x^{\mu}$ and $n$, and 
\begin{align}
\left.D^2\right|_{A^E_{\mu},A^E_5,c_A}
&=-\partial^{2}_{E}+\left(\frac{ n}{L_5}\right)^2+m_\cA^2,\notag\\
\left.D^2\right|_{\tilde \theta}
&=
-\partial^2_{E}+
\left(
 \left(\frac{n}{L_5}\right)^2
 +m_\cA^2
 \right) 
\left(1+\frac{8v_4}{m_\cA^2}(\cA^{c}_5)^2
\right) .
\end{align}
The effective potential is given as
\begin{align}
V_g(A)
=
\sum_{n=-\infty}^{\infty}
\int
\frac{d^4p_E}{(2\pi)^4}
&
\left[\frac{3}{2}
{\rm ln}
\left\{p_E^2+\left(\frac{n}{L_5}\right)^2
+m_\cA^2
\right\}
\right.
\notag\\
&\left.
+\frac12{\rm ln}
\left\{
p_E^2
+
\left(
\left(\frac{n}{L_5}\right)^2
+m_\cA^2
\right)
\left(1+\frac{8v_4}{m_\cA^2}(\cA^{c}_5)^2\right)
\right\}
\right].
\end{align}
The $\zeta$ function regularization yields the following result:
\begin{align}
V_g(A)
&=
-\frac{1}{4\pi^2(2\pi L_5)^2}\left[3+\left(1-\frac{\lambda}{3}\frac{A^2}{m^2_\cA}\right)^{2}\right]\sum_{k=1}^{\infty}\frac{m_\cA^2}{k^3}
\left(1+3(kz)^{-1}+3(kz)^{-2}\right)
e^{-kz},\nn \\
&
\label{gaugepot}
\end{align}
where
\begin{equation}
z:= \sqrt{m^2_\cA (2 \pi L_5)^2},
\end{equation}
and (\ref{amass1}) is rewritten as
\begin{equation}
m^2_\cA
=
-2v_2+\frac{4v_4}{(2\pi L_5)}A^2
= m^2
-\frac{\lambda}{6}A^2, 
\qquad m^2,\lambda>0.
\label{amass2}
\end{equation}
Adding classical potential (\ref{potex})
and the one-loop contributions (\ref{appVf}) and (\ref{gaugepot}), 
we obtain the following effective potential for $A$ and $B$:
\begin{align}
V_{1-loop}(A,B)
=&
V_{cl}(A)+V_f(A,B)+V_g(A)\notag\\
=&
\frac{m^2}{2}A^2-\frac{\lambda}{4!}A^4
+\frac{3}{\pi^2 (2\pi L_5)^4}
\sum_{n=1}^{\infty}\frac{1}{n^5}
\cos \left[n\left(\frac{A}{f_A}-\frac{B}{f_B}\right)\right]\notag\\
&-\frac{1}{4\pi^2(2\pi L_5)^2}
\left[3+\left(1-\frac{\lambda}{3}\frac{A^2}{m^2_\cA}\right)^{2}\right]
\sum_{k=1}^{\infty}\frac{m_\cA^2}{k^3}
\left(1+3(kz)^{-1}+3(kz)^{-2}\right)
e^{-kz}.
\notag\\
\label{gaugepot2}
\end{align}

\subsection{The comparison between $V_g(A)$ with $V_{cl}(A)$}

In what follows,
we examine whether and when
the contributions of $V_g(A)$
to the energy density and spectral index
are sub-leading compared with $V_{cl}(A)$.
For this purpose, it is convenient to change the variable 
from $A$ to $\phi \sim \frac{f_B}{f_A} A$.
To estimate $V_g(\phi)$,
taking only $k=1$ term in (\ref{gaugepot2})
is a good approximation:
\begin{align}
V_g(\phi_*)
\sim&-\frac{m^2(1-2c\phi_*^2)}{4\pi^2(2\pi L_5)^2}
\left[
3+\left(1-\frac{4c\phi_*^2}{1-2c\phi_*^2}\right)^{2}
\right]
\left(
1+3z^{-1}+3z^{-2}
\right)
e^{-z} \nn\\
 &+ \frac{3}{\pi^2(2\pi L_5)^4},
\label{Vgphi}
\end{align}
with
\begin{equation}
z = (2 \pi L_5) m (1-2c\phi_\ast^2)^{1/2}.
\end{equation}
In the above we have added the constant term
so that $V_g(0)=0$ is satisfied, 
in order to tune the cosmological constant.
Since we are mostly interested in the case
$|V_g(\phi)| \ll V_{cl}(\phi)$, 
we subtracted $V_g(0)$ instead of the energy density 
at the minimum of the total potential for simplicity.
To estimate (\ref{Vgphi}),
we first observe from Fig.~\ref{fig:cm} that
$m_{eff} \lesssim 7\times 10^{-6}$.
Thus if we take $f_B/f_A \sim 10-30$,
$m=\frac{f_B}{f_A}m_{eff}\lesssim 7\times 10^{-5}- 2\times 10^{-4}$.
On the other hand, from Fig.~\ref{fig:L} 
we observe that
$2\pi L_5 \lesssim 10^3$,
thus $m (2\pi L_5) \lesssim 2 \times 10^{-1}$.
Taking the leading term 
in the power series expansion
in $m(2\pi L_5)$, we obtain
\begin{align}
V_g (\phi) 
&\sim
-\frac{3}{4\pi^2(2\pi L_5)^4}
\left\{
3 +
\left(
1 - \frac{4c\phi^2}{1-2c\phi^2}
\right)^2
\right\} 
+
\frac{3}{\pi^2(2\pi L_5)^4}
\nn\\
&= 
\frac{3}{\pi^2(2\pi L_5)^4}
\left. v_{g} (x) \right|_{x=\sqrt{c}\phi} ,
\label{Vgvg}
\end{align}
where
\begin{equation}
v_g(x) 
=
\frac{1}{4}
\left\{
3 +
\left(
1 - \frac{4x^2}{1-2x^2}
\right)^2
\right\}
-1 .
\label{vgx}
\end{equation}
To compare (\ref{Vgvg}) with $V_{cl}(\phi)$,
we rewrite $V_{cl}(\phi)$ as
\begin{equation}
V_{cl}(\phi) 
= \frac{m_{eff}^2 \phi^2}{2} 
\left(
1 - c \phi^2
\right)\nn\\
=
\frac{m_{eff}^2}{2c} \left. v_{cl} (x) \right|_{x = \sqrt{c}\phi} ,
\end{equation}
where
\begin{equation}
v_{cl}(x) 
=
x^2 (1 - x^2). 
\label{vclx}
\end{equation}
Near the observationally 
preferable point 
$c = 0.001$ and $N_\ast =60$,
$\phi_\ast \lesssim 15$ and thus
$\sqrt{c}\phi_\ast \lesssim 0.5$.
In the domain $0 \leq x < 0.5$
the functions 
$v_{g}(x)$, 
$v_{cl}(x)$ and their derivatives are roughly of order one,
thus for a crude comparison between $V_g(\phi)$ and $V_{cl}(\phi)$
we can compare the coefficients in front of these functions,
$
{3}/{\pi^2(2\pi L_5)^4}
$
and
$
{m_{eff}^2}/{2c}
$.
When $c=0.001$ and $N_\ast =60$,
$
{m_{eff}^2}/{2c} \sim 6 \times 10^{-9}
$,
thus
the contributions of $V_g(\phi)$
to the energy density and the spectral index
compared with those of $V_{cl}(\phi)$
are sub-leading if
\begin{equation}
\frac{3}{\pi^2(2\pi L_5)^4} 
\lesssim 
6 \times 10^{-9},
\label{sub}
\end{equation}
or equivalently
\begin{equation}
2\pi L_5 \gtrsim 1 \times 10^2 .
\label{2piL5}
\end{equation}
Since the five-dimensional gauge theory is not
renormalizable and is regarded as an effective field theory,
it is natural that
the compactification radius $L_5$
is not too close to the Planck scale.
Therefore, (\ref{2piL5}) is a natural condition to impose
and we have assumed this in the main body.

\bibliography{XDI2ref}
\bibliographystyle{JHEP}
\end{document}